\documentclass[%
 aip,
 jmp,%
 amsmath,amssymb,
]{revtex4-1}

\draft

\usepackage{graphicx}% Include figure files
\usepackage{dcolumn}% Align table columns on decimal point
\usepackage{bm}% bold math

\usepackage{framed}
\usepackage{siunitx} % to get proper distances between num and unit 
\usepackage{xspace}
\usepackage{xargs} % Use more than one optional parameter in a new commands
\usepackage{multirow}
\usepackage{nicefrac}
\usepackage[colorinlistoftodos,prependcaption,textsize=tiny]{todonotes}
\newcommandx{\unsure}[2][1=]{\todo[linecolor=red,backgroundcolor=red!25,bordercolor=red,#1]{#2}}
\newcommandx{\change}[2][1=]{\todo[linecolor=blue,backgroundcolor=blue!25,bordercolor=blue,#1]{#2}}
\newcommandx{\info}[2][1=]{\todo[linecolor=OliveGreen,backgroundcolor=OliveGreen!25,bordercolor=OliveGreen,#1]{#2}}
\newcommandx{\improvement}[2][1=]{\todo[linecolor=Plum,backgroundcolor=Plum!25,bordercolor=Plum,#1]{#2}}
\newcommandx{\thiswillnotshow}[2][1=]{\todo[disable,#1]{#2}}

\newcommand{\sr}{Sr$_\text{Hf}^{-2}$\xspace}
\newcommand{\srvo}{$\left[\text{Sr}_\text{Hf}V_\text{O}\right]^0$\xspace}
\newcommand{\vo}{V$_\text{O}^{+2}$\xspace}

\DeclareSIUnit[]\formulaunit{f.u.}

\usepackage{hyperref}% add hypertext capabilities
\usepackage[mathlines]{lineno}% Enable numbering of text and display math
\begin{document}

\preprint{APL}

\title{The impact of charge compensated and uncompensated strontium defects on the stabilization of the ferroelectric phase in HfO2}

% Force line breaks with \\
%\thanks{A footnote to the article title}%

\author{Robin Materlik}
%\email{robin.materlik@hm.edu}
\affiliation{Department of Applied Sciences and Mechatronics, Munich University of Applied Sciences, Lothstr. 34, 80335 Munich, Germany}%
\affiliation{These two authors contributed equally to this work.}

\author{Christopher K\"unneth}%
%\email{kuenneth@hm.edu}
\affiliation{Department of Applied Sciences and Mechatronics, Munich University of Applied Sciences, Lothstr. 34, 80335 Munich, Germany}%
\affiliation{These two authors contributed equally to this work.}

\author{Thomas Mikolajick}
\affiliation{NaMLab gGmbH, Noethnitzer Strasse 64, 01187 Dresden, Germany}%
\affiliation{Chair of Nanoelectronic Materials, Technische Universit\"at Dresden, Noethnitzer Strasse 64, 01187 Dresden, Germany}%

\author{Alfred Kersch}
\email{alfred.kersch@hm.edu}
\affiliation{Department of Applied Sciences and Mechatronics, Munich University of Applied Sciences, Lothstr. 34, 80335 Munich, Germany}%

\date{\today}% It is always \today, today,
             %  but any date may be explicitly specified

\begin{abstract}
Different dopants with their specific dopant concentration can be utilized to produce ferroelectric HfO$_2$ thin films. In this work it is explored for the example of Sr in a comprehensive first-principles study. Density functional calculations reveal structure, formation energy and total energy of the Sr related defects in HfO$_2$. We found the charge compensated defect including an associated oxygen vacancy Sr$_\text{Hf}$V$_\text{O}$ to strongly favour the non-ferroelectric, tetragonal P4$_\text{2}$/mnc phase energetically. In contrast, the uncompensated defect without oxygen vacancy Sr$_\text{Hf}$ favours the ferroelectric, orthorhombic Pca2$_\text{1}$ phase. According to the formation energy the uncompensated defect can form easily under oxygen rich conditions in the production process. Low oxygen partial pressure existing over the lifetime promotes the loss of oxygen leading to V$_\text{O}$ and, thus, the destabilization of the ferroelectric, orthorhombic Pca2$_\text{1}$ phase accompanied by an increase of the leakage current. This study attempts to fundamentally explain the stabilization of the ferroelectric, orthorhombic Pca2$_\text{1}$ phase by doping.
\vspace{6em}
\begin{framed}
	This article may be downloaded for personal use only. Any other use requires prior permission of the author and AIP Publishing. The following article appeared in Appl. Phys. Lett., vol. 111, no. 8, p. 82902, 2017 and may be found at http://dx.doi.org/10.1063/1.4993110.	
\end{framed}

\end{abstract}

%\pacs{Valid PACS appear here}% PACS, the Physics and Astronomy
                              % Classification Scheme.
\maketitle

Polycrystalline HfO$_\text{2}$ thin films produced by Atomic Layer Deposition (ALD) or Chemical Solution Deposition (CSD) can exhibit ferroelectric properties if they are appropriately doped \cite{Boscke2011, Muller2012, Park2013, Lomenzo2014, Shimizu2016, Starschich2017, AdvancedMaterials, Schroder2014, Olsen2012}. An orthorhombic, non-centrosymmetric phase (Pca2$_\text{1}$) has been proposed as the source of these properties which has since been confirmed by electron diffraction study \cite{Sang2015}. Furthermore, another theoretically proposed ferroelectric Pmn2$_\text{1}$ phase has been ruled out by the same study and is therefore not included in this work. Pure HfO$_\text{2}$ occurs naturally in a monoclinic (P2$_\text{1}$/c) phase. With increasing temperature a transformation into the tetragonal (P4$_\text{2}$/mnc) and then the cubic (Fm$\overline{3}$m) phase occurs \cite{Wang1992} avoiding the orthorhombic phase. Different Density Functional Theory (DFT) studies consistently calculate the total energy of the orthorhombic phase as second most stable after the monoclinic phase and are able to reproduce the thermally driven phase transformation \cite{Huan2014, Materlik2015} giving credibility to the used density functionals.

To explain the occurrence of the ferroelectric phenomena, factors favouring the orthorhombic phase have been proposed \cite{DeLillo2015, Materlik2015, Batra2016} including entropy contribution, surface or interface energy, stress, and doping. Surface or interface energy stems from the large surface to volume ratio of the individual crystals in the polycristalline HfO$_\text{2}$ thin films \cite{Materlik2015, Kuenneth2017} with grain sizes typically in the range of the film thickness (\SIrange{5}{30}{nm}) \cite{Polakowski2015, Park2015, HyukPark2013, Kim2014}. It explains the generally observed decrease or disappearance of the ferroelectric properties with increasing film thickness\cite{Hoffmann2015}. For the case of Hf$_\text{1-x}$Zr$_\text{x}$O$_\text{2}$, at $x = 0.5$, surface energy or interface energy has been found to be sufficient to explain stability of the orthorhombic phase \cite{Materlik2015, Kuenneth2017}. For thin films based on pure HfO$_\text{2}$ surface or interface energy is insufficient except for the case of very small grains \cite{Polakowski2015}.

In such thin films further stabilization by appropriate doping is required \cite{AdvancedMaterials, Mueller2012, Olsen2012, Schroder2014, Mueller2012b}. For the case of Sr doping, ferroelectricity was observed in a \SI{10}{nm} film between 1.7 and \SI{7.9}{mol\%\ SrO} content with the maximum polarization observed at around \SI{3.4}{mol\%\ SrO} \cite{Schenk2013}. The effect of doping on HfO$_\text{2}$ phases has been investigated in earlier works \cite{Choong-KiLee2008, Fischer2008} but the Pca2$_\text{1}$ as well as II-valent dopants were not included in the study. The authors found stabilization of the tetragonal phase by IV-valent and stabilization of the cubic phase by III-valent dopants. Due to its II-valent nature, it is expected that each Sr dopant atom is accompanied by an oxygen vacancy for charge compensation. Furthermore, due to opposite charges, the \sr and \vo defect should strongly attract each other leading to \srvo similar to the case of $\text{Mg}_\text{Hf}^{-2}$ or $\text{Ba}_\text{Hf}^{-2}$ doping investigated in \cite{Umezawa2008, Umezawa2009}. However, the defect concentration created during the manufacturing process is not explicit known and strongly depends on the chemical potential of the defects. In this work the defect notation of Freysoldt et al. is used \cite{Freysoldt}.

To propose a consistent scenario for the ferroelectric stability of a Sr doped HfO$_\text{2}$ thin film, we determined total energy and defect formation energy for various defects in monoclinic, orthorhombic, tetragonal and cubic HfO$_\text{2}$ from first principle calculations. These defects include single oxygen vacancies V$_\text{O}$$^{q}$ with the charges $q = 0,+1,+2$, Sr substituted for Hf with Sr$_\text{Hf}$$^{q}$ ($q = 0, -1, -2$) as well as the compensated defect [Sr$_\text{Hf}$ V$_\text{O}$]$^{q}$($q= 0, -1, -2$). Oxygen vacancies were placed on the eight next neighboring oxygen sites of a given Sr or Hf atom excluding structural equivalent positions. All shown results always depict the energetically most favourable position. Placing one defect in a 96 or 48 atomic super cell corresponds to a concentration of \SI{3.125}{\formulaunit\%} (= 1 defect/ 32 formula units) and \SI{6.25}{\formulaunit\%} (= 1 defect/ 16 formula units) respectively.

DFT calculations were performed using the Local Density Approximation (LDA) and Projector Augmented Wave (PAW) \cite{Blochel1994} Pseudo Potentials (PP) from the GBRV library \cite{GBRV, Garrity2014} with the ABINIT code \cite{Gonze2009,Gonze2016,Torrent2008}. Several LDA calculations were repeated with the all electron code FHI-AIMS \cite{Blum2009} based upon numeric, atom-centered orbitals of type tight with first and second tier enabled. In the remainder of this work we will refer to those two methods as plane waves (PW) and numerical orbitals (NO), respectively. The stopping criteria for the electronic convergence was a force criteria of $10^{-6}$ Hartree/Bohr (PW) and $10^{-4}$ eV/\AA (NO). The stopping criteria for the structural convergence was a force criteria of $10^{-5}$ Hartree/Bohr (PW) and $10^{-3}$ eV/\AA (NO).
Charged and neutral defect calculations in monoclinic, tetragonal, cubic, and orthorhombic HfO$_\text{2}$ were performed with 96 atomic super cells using a $2 \times 2 \times 2$ Monkhorst-Pack k-point set, a plane wave cut off of 18 Ha and a PAW cut off of 22 Ha in accordance with a convergence study. Charge neutral 48 atomic super cells with a $2 \times 4 \times 2$ k-point grid were used to determine the phase stability at \SI{6.25}{\formulaunit\%} defect concentration.

The defect formation energies $E_\text{f}$ were calculated as 
\begin{eqnarray}
&&E_\text{f}\left(X, q\right) = U\left(X, q\right)-U\left(\text{pure}\right)-\sum_{i}{ n_{i}\mu_{i} }\nonumber\\
&&+q\left(\epsilon_{\text{F}}+\epsilon_{\text{VB}}\left(\text{pure}\right) + \Delta V\left(X, 0\right) \right)+E_{\text{Corr}}\left(X, q\right) 
\label{eq:formation}
\end{eqnarray}
using the DFT total energies $U$ of both HfO$_\text{2}$ without and with a defect $X\in \left\{ \text{Sr}_\text{Hf}^{q},  \left[\text{Sr}_\text{Hf}\text{V}_\text{O}\right]^{q}, \text{V}_\text{O}^{q}\right\}$ and charge $q$. The chemical potential and number of defect atoms of each species is given by $\mu_{i}$ and $n_i$ respectively. The Fermi energy is $\epsilon_\text{F}$ and the valence band edge is $\epsilon_\text{VB}$. A charge correction $E_\text{Corr}$ with the scaling law\cite{MakovPayne1995} $E_\text{f} \sim \nicefrac{a}{L} + c$ using a 324 atomic super cell and a potential alignment $\Delta V$ was applied. $a$ and $c$ are fit parameters and $L$ is the size of the super cell. The chemical potential of Hf was set to the total energy of hcp Hf and of Sr was calculated by the equilibrium condition $\mu_\text{Sr}$ = $\mu_\text{SrO}$ - $\mu_\text{O}$. For the chemical potential of oxygen two cases are considered: oxygen rich and oxygen deficient \cite{Tang2010, Lyons2011}. In the oxygen rich case  $\mu_\text{O}$ is set to $\nicefrac{\mu_ {\text{O}_{2}}}{2}$. Ferroelectric HfO$_\text{2}$ is often deposited on TiN electrodes \cite{Pesic2016, Kim2014, Polakowski2015, Park2015, HyukPark2013, Sang2015, Boscke2011, Muller2012, Schenk2013} which can exist in a partially oxidized state. The oxygen chemical potential $\mu_\text{O}$ for the deficient conditions uses oxygen precipitation into anatase TiO$_\text{2}$. In similar studies \cite{Umezawa2009}, precipitation into SiO$_\text{2}$ has been used adapting to a Si substrate. Both assumptions, however, lead to very similar formation enthalpies. We therefore calculate $\mu_\text{O}=\nicefrac{\left(\mu_{\text{TiO}_{2}}-\mu_\text{Ti}\right)}{2}$ for the oxygen deficient case.

\begin{figure}
	\includegraphics{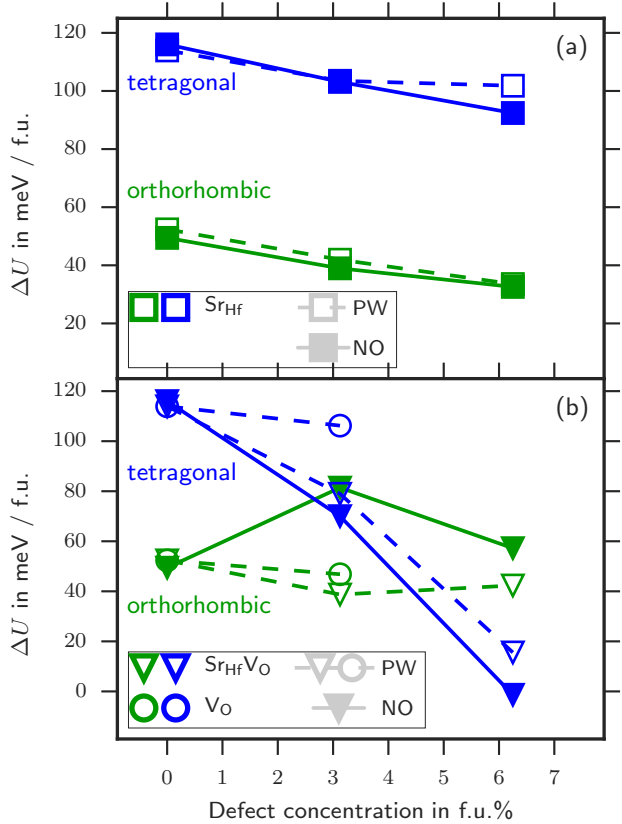}••••••••
	\caption{ Defect concentration dependent energy difference $\Delta U = U\left(o \| t\right) - U\left(m\right)$ to the monoclinic phase for the PW (empty symbols, dashed line) and NO (full symbols, continuous line) methodology for the tetragonal (blue) and orthorhombic (green) phase. The different defects are indicated by symbols. (a) shows the vacancy free defects Sr$_\text{Hf}$ (squares) and (b) shows the vacancy related defects V$_\text{O}$ (triangles) and Sr$_\text{Hf}$V$_\text{O}$ (circles).}
	\label{fig:defect_concentrations}
\end{figure}

The main result from this paper is the connection between the phase stability of defective HfO$_\text{2}$ and the conditions under which the defective material can form.
FIG. \ref{fig:defect_concentrations} (a) shows the total energy difference $\Delta U$ to the monoclinic phase for the Sr$_\text{Hf}$ defect as a function of the Sr concentration. Both the orthorhombic and tetragonal phase are depicted calculated with PW and NO. The cubic phase turned out to be unstable and is therefore not shown here.

The defect free orthorhombic phase has a $\Delta U$ of \SI{53}{meV} (PW) and \SI{49}{meV} (NO) while the tetragonal phase has \SI{115}{meV} (PW) or \SI{114}{meV} (NO). The Sr$_\text{Hf}$ defects lead to a decrease of $\Delta U$ of about \SI{20}{meV} for \SI{6}{Sr-\formulaunit\% } which is roughly the same for both the tetragonal and orthorhombic phase. Therefore, the defect contributes to the stabilization but not sufficiently to fully stabilize the orthorhombic phase on its own. However, according to previous works, in Hf$_\text{1-x}$Zr$_\text{x}$O$_\text{2}$ \cite{Materlik2015, Kuenneth2017, Garvie1965} the surface or interface energy of grains can decrease the energy of the tetragonal and orthorhombic phase below the monoclinic phase and, thus, suppress the formation of the monoclinic phase. The surface or interface energy for the tetragonal and orthorhombic phase are expected to be very similar. Proposing a surface or interface energy penalty for the monoclinic phase is difficult in this case since the issue has not been investigated for doped HfO$_\text{2}$ so far. In Hf$_\text{1-x}$Zr$_\text{x}$O$_\text{2}$ considering Zr as a dopant, a typical energy penalty of about \SI{20}{meV} (for typical grains of \SI{10}{nm} diameter in a \SI{10}{nm} film) was found for HfO$_\text{2}$ linearly increasing to about \SI{60}{meV} for ZrO$_2$. At the same time the interface energies increased from \SIrange{174}{490}{mJ/m^2}~~\cite{Kuenneth2017}. There is another argument in favour of a significant increase of the energy penalty for the monoclinic phase with doping. The authors \cite{Kuenneth2017} indentified the energy penalty with the energy of the tetragonal/monoclinic interface observed by Grimley \cite{Grimley2016}. An interface energy, however, is expected to depend sensitively on doping. Altogether, we expect a surface or interface related energy penalty for the monoclinic phase starting at around \SI{30}{meV} for pure HfO$_\text{2}$ and increasing significantly with doping. We therefore expect $\Delta U$ of the orthorhombic phase to become negative for some Sr concentrations and the film to become ferroelectric. Important for the ferroelectric stabilization is that the orthorhombic phase turns always out to be more favourable than the tetragonal phase.

This is not the case for the compensated defect \srvo as shown in FIG. \ref{fig:defect_concentrations} (b). The change in $\Delta U$ is much larger for the tetragonal phase than for the orthorhombic phase. Above a threshold of \SIrange{2}{3}{\formulaunit\%} (NO) or \SI{5}{\formulaunit\%} (PW) the material looses ferroelectricity and the tetragonal phase replaces the orthorhombic phase as the most favorable. This would severely limit the dopant concentration range in which ferroelectric properties can be observed and is therefore in conflict with the experimentally observed range for ferroelectricity of \SIrange{1.7}{7.9} {mol\%} dopant concentration\cite{Schenk2013, Schroder2014}.

This leads to the question, whether the Sr$_\text{Hf}$ is indeed always compensated with an oxygen vacancy V$_\text{O}$ as stoichiometry suggests. An estimation of the vacancy concentration results from an electrical measurement of the leakage current in Sr doped Hf by Pe\v{s}i\'{c} et al. \cite{Pesic2016}, who extracted a vacancy concentration of \SI{5e19}{cm^{-3}} which is significantly less than required to pair every Sr atom (\SI{1.4e21}{cm^{-3}} for \SI{5}{\formulaunit\%}) with a vacancy. Crucial for the question, whether Sr$_\text{Hf}$ or Sr$_\text{Hf}$V$_\text{O}$ should be expected is the formation energy as a function of the oxygen chemical potential and a kinetic process creating the defect \cite{McIntyre}.

\begin{figure}
	\includegraphics{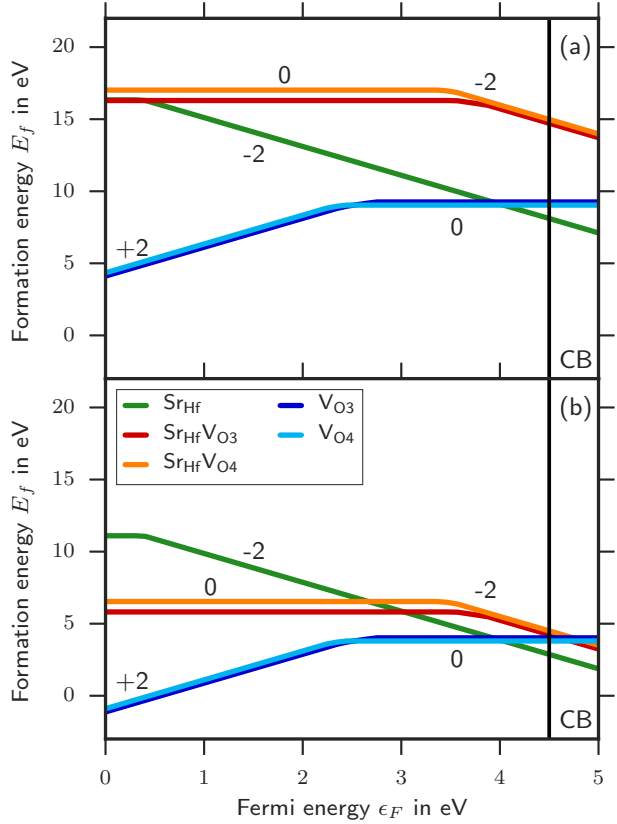}
	\caption{Figure 2 shows the formation energies of the orthorhombic phase for (a) oxygen rich and (b) oxygen deficient conditions, respectively. The formation energies are calculated by EQ. \ref{eq:formation} and the charge states are indicated by numbers. The values are not scaled to the experimental band gap and CB marks the LDA-DFT calculated conduction band.}
	\label{fig:formation_energies}
\end{figure}

FIG. \ref{fig:formation_energies} (a) shows the formation energies under oxygen rich conditions for the orthorhombic phase and for oxygen in the III-valent and IV-valent position. The formation energy does not differ very much from the monoclinic phase (not shown here). The LDA band gap for the orthorhombic phase was found to be \SI{4.41}{eV} (\SI{3.98}{eV} for the monoclinic and \SI{4.56}{eV} for the tetragonal). The individual formation energy of charged \sr and \vo defects is lower than the formation energy of the combined charge neutral \srvo defect. This might lead to a separated creation of \sr and \vo. However, since vacancies are very mobile, the positively charged vacancies \vo combine with the negatively charged \sr creating \srvo with an energy release of \SI{2.36}{eV}. Under oxygen rich conditions, few Sr$_\text{Hf}$V$_\text{O}$ are expected in the end except close to the interface where some oxygen loss towards the electrode has to be expected. As a result, a film with substitutional Sr$_\text{Hf}$ defects and few compensated defects is expected, but at the electrode interface a significant amount of compensated Sr$_\text{Hf}$V$_\text{O}$ defects is possible which may stabilize a tetragonal interlayer \cite{Grimley2016} and may be a prerequisite of the energy penalty to suppress the monoclinic phase. This would support the assumptions made by Pe\v{s}i\'{c} \cite{Pesic2016}. The acceptor doping without charge compensation achieved under oxygen rich conditions is often desired to improve electric isolation since the negative space charge increases the band offset to the electrode.

During the life time of a ferroelectric HfO$_\text{2}$ stack, the external oxygen partial pressure is defined by the oxidized electrodes. FIG. \ref{fig:formation_energies} (b) shows the formation energy under such oxygen deficient conditions. As there is no new Sr-source, only vacancies can be created possibly due to field cycling. Since the energy of \srvo is lower than the sum of \vo and \sr, these vacancies will recombine quickly with the already present substitutional Sr defects leading to a charge compensation. The concentration of \sr will decrease and that of \srvo will increase. The implication on the phase stability is a gradual degradation of the orthorhombic phase content accompanied by a decrease of the remanent polarization. A further implication concerning the electron transport is that the charge transition level of a deep defect state promotes trap assisted tunneling (TAT). The related charge transition levels $\epsilon\left(0/-1\right)=$ \SI{3.63}{eV} and $\epsilon\left(-1/-2\right) =$ \SI{3.92}{eV} close to the conduction band release electrons which modify the space charge and contribute to TAT. Therefore, a moderate increase of leakage current with time would be expected indicating an increase of charge compensated defects. 
As the creation of Sr$_\text{Hf}$V$_\text{O}$ under oxygen deficient conditions is preferred, the concentration of V$_\text{O}$ will stay on a relatively low level and constant over time. However, the V$_\text{O}$ defects with charge transition levels at $\epsilon\left(+2/+1\right)=$ \SI{2.41}{eV} and $\epsilon\left(+1/0\right)=$ \SI{2.81}{eV} are about \SI{2}{eV} below the conduction band and, therefore, can be occupied by tunneling electrons promoting leakage current. 

A last argument explains why the Sr$_\text{Hf}$ defect favors the orthorhombic and Sr$_\text{Hf}$V$_\text{O}$ defect the tetragonal phase in total energy. The cause for the stabilization of the orthorhombic and tetragonal phase by Sr$_\text{Hf}$ defects can be found in the bond length of the Sr atom to its neighboring oxygen atoms. Calculations of SrO and SrO$_\text{2}$ show a bond length between 2.53 and \SI{2.60}{\AA}, respectively. In undoped HfO$_\text{2}$ the average bond length is \SI{2.12}{\AA} for the monoclinic and orthorhombic phase and \SI{2.17}{\AA} for the tetragonal phase. Substituting a Sr atom on a Hf site, the bond length increases to only \SI{2.35}{\AA} for the monoclinic phase but to 2.37 {\AA} for the orthorhombic and tetragonal phase. Sr in monoclinic HfO$_\text{2}$ is therefore energetically more unfavourable than in the orthorhombic or tetragonal phase, therefore the energy difference to the monoclinic phase decreases with doping. Introducing vacancies, the monoclinic average bond length increases to \SI{2.38}{\AA}, but the tetragonal value of \SI{2.47}{\AA} almost matches the value of SrO and is accompanied by the significant decrease in total energy difference, see FIG. \ref{fig:defect_concentrations}.

In summary a mechanism is proposed, based on first-principles DFT calculations, to explain the influence of Sr doping on the phase stability in HfO$_\text{2}$. The tetragonal phase is strongly preferred by the incorporation of the Sr$_\text{Hf}$V$_\text{O}$ defects while the Sr$_\text{Hf}$ allows for the stabilization of the ferroelectric orthorhombic phase. The uncompensated defect can form in sufficiently oxygen rich environments, which might exist during the production process. The loss of oxygen during field cycling may increase the charge compensation which promotes the phase transformation into other HfO$_\text{2}$ polymorphs. This contributes to the fatigue behavior. The proposed mechanism has the potential to describe the action of other dopants on the ferroelectric phase in HfO$_\text{2}$ if appropriately adapted and expanded.  

%\begin{acknowledgments}
The author wants to thank U. Schr\"{o}der, T.Schenk, Min Hyuk Park from NamLab/SNU, and U. B\"{o}ttger and S. Starschich from RWTH Aachen for discussions. The German Research Foundation (Deutsche Forschungsgemeinschaft) is acknowledged for funding this research in the frame of the project “Inferox” (Project No. MI 1247/11–1). The authors gratefully acknowledge the Gauss Centre for Supercomputing e.V.  (www.gauss-centre.eu)  for  funding  this  project  by providing  computing  time  on  the  GCS  Supercomputer SuperMUC  at  Leibniz Supercomputing  Center  (LRZ, www.lrz.de).
%\end{acknowledgments}

\bibliography{bib}% Produces the bibliography via BibTeX.

\end{document}